\documentstyle[12pt,aaspp4]{article}
\begin{document}

% Next 5 lines define \simless and \simgreat: "less than or approximately
% equal to" and "greater than or approximately equal to".
\newbox\grsign \setbox\grsign=\hbox{$>$} \newdimen\grdimen \grdimen=\ht\grsign
\newbox\simlessbox \newbox\simgreatbox
\setbox\simgreatbox=\hbox{\raise.5ex\hbox{$>$}\llap
     {\lower.5ex\hbox{$\sim$}}}\ht1=\grdimen\dp1=0pt
\setbox\simlessbox=\hbox{\raise.5ex\hbox{$<$}\llap
     {\lower.5ex\hbox{$\sim$}}}\ht2=\grdimen\dp2=0pt
\def\simgreat{\mathrel{\copy\simgreatbox}}
\def\simless{\mathrel{\copy\simlessbox}}

\title{NEAR-INFRARED SPECTRAL FEATURES IN SINGLE-AGED STELLAR POPULATIONS}

\author{\bf R. P. Schiavon\altaffilmark{1,2}, 
B. Barbuy\altaffilmark{3}, G. Bruzual A.\altaffilmark{4}}

\altaffiltext{1}{Observat\'orio Nacional/CNPq, Departamento de Astronomia,
R. Gen. Jos\'e Cristino 77, 20921-400 Rio de Janeiro, Brazil. 
Email: ripisc@ucolick.org}

\altaffiltext{2}{Present address: UCO/Lick Observatory, University of
California, Santa Cruz, CA, 95064}

\altaffiltext{3}{Universidade de S\~ao Paulo, IAG, Departamento de Astronomia,
C.P. 3386, S\~ao Paulo 01060-970, Brazil.
Email: barbuy@orion.iagusp.usp.br}

\altaffiltext{4}{Centro de Investigaciones de Astronom{\'\i}a (CIDA), Apartado
Postal 264, M\'erida 5101-A, Venezuela. Email: bruzual@cida.ve} 

\slugcomment{Submitted to The Astrophysical Journal Letters}
\slugcomment{Send proofs to: R. P. Schiavon (Address 2)}

\begin{abstract}
Synthetic spectra for single-aged stellar populations of metallicities
[M/H] = -0.5, 0.0 and +0.5, ages = 3 to 17 Gyrs, and initial
mass function indices $x$ = 0.1 to 2.0 were built in the wavelength
range  $\lambda\lambda$ 6000-10200 {\rm \AA}.

For such we have employed the grid of synthetic spectra described
in Schiavon \& Barbuy (1999), computed for the stellar
parameters 2500 $\leq$ T$_{\rm  eff}$ $\leq$ 6000 K, 
-0.5 $\leq$ log g $\leq$ 5.0, [M/H] = -0.5, 0.0 and +0.5,
and [$\alpha$/Fe] = 0.0, together with
the isochrones by Bertelli
et al. (1994) and Baraffe et al. (1998).

The behavior of the features NaI$\lambda$8190,
CaII$\lambda$8662, TiO$\lambda$6600
and FeH$\lambda$9900 {\rm \AA} in the integrated spectra of
 single-stellar populations
were studied in terms of metallicity, initial mass function
and age variations.

The main conclusions are that the NaI doublet is an IMF-sensitive
feature, which is however sensitive also to metallicity and age,
whereas TiO, CaII and FeH are
very sensitive to metallicity and essentially insensitive to IMF
and age. 

\end{abstract}


\section{INTRODUCTION}

The analysis of near-infrared (NIR) features in the spectral region
$\lambda\lambda$ 6000-10000 {\rm \AA} in the integrated light of
globular clusters and galaxies has become of common use
 to infer the parameters of their stellar populations. The NIR
spectral region is specially interesting for stellar population studies
because M stars (giants and dwarfs)
give their maximum contribution to the integrated spectrum
of galaxies and clusters in these wavelengths. 
As an example, NIR line indices
which are sensitive to the surface gravity of M stars can be used to
constrain the low mass end of the mass function.
Also, NIR indices are suitable for deriving metallicities 
because of their remarkable insensitivity to age. 
The metallicity calibration of globular clusters
using the CaII triplet has been carried out by Bica et al. (1998) for
globular clusters in the Galactic bulge, and by Olszewski et al. (1991)
for clusters in the Magellanic Clouds.

Long standing problems with the use of 
IMF-sensitive NIR features in composite systems
were discussed in
Whitford (1977), Cohen (1978, 1979), Faber \& French (1980), 
Alloin \& Bica (1989), Xu et al. (1989), Boroson \& Thompson (1991), 
Delisle \& Hardy (1992) and Couture \& Hardy (1993), which studied
the NaI, CaII and FeH indices in integrated spectra of galaxies and
clusters.

In spite of the above mentioned efforts, the strong spectral features
which appear in the integrated NIR spectrum of globular
clusters and normal galaxies have been far less exploited in stellar
population studies than the indices in the optical region, such as
Mg$_2$, Fe5270, and Fe5335 (Burstein et al. 1984; Worthey 1994;
Trager et al. 1998). This fact is probably due to two main reasons: the
NIR features show a more complex behavior than the `green' features,
and the NIR region presents numerous telluric lines which require a
proper and careful subtraction.

In this context, spectrum synthesis based on  model stellar
photospheres and comprehensive line lists can play a key r\^ole in
disentangling the degenerate behaviour of line indices as a function
of metallicity, effective temperature and surface gravity. In an
attempt to improve our understanding of the NIR features, Schiavon \&
Barbuy (1999, hereafter SB99) have built an extensive grid of synthetic
spectra, based on state-of-the-art model photospheres and molecular and
atomic line lists.  The NaI `doublet' at $\lambda$8190{\rm \AA} and the
FeH Wing-Ford band (WFB) at 9900{\rm \AA} (Wing \& Ford 1969) were
studied as a function of stellar photospheric parameters by
Schiavon et al. (1997a, b), whereas the behavior of the TiO bands and
the CaII triplet were described in Milone \& Barbuy (1994), SB99 and
Erderlyi-Mendes \& Barbuy (1991).

In this paper we use the grid of high resolution synthetic stellar
spectra presented in SB99, to build integrated spectra of single-aged
stellar populations (SSPs) in the wavelength range $\lambda\lambda$
6000-10200 {\rm \AA}.  Previous work with similar aims was presented by
Garcia-Vargas et al. (1998) and Milone et al. (1995) where equivalent
widths  of the CaII triplet and TiO respectively  were given.

These models are a useful guide in the interpretation of the integrated
spectrum of early-type galaxies, which in the NIR are dominated by
M-type stars, and for a better understanding of the age-metallicity
degeneracy of old and intermediate age stellar populations.

In \S2 we present basic information on the spectrum synthesis of
individual stellar models and on the other ingredients required to
build the SSP models. In \S3 we discuss the behavior of selected NIR
features as a function of SSP parameters: metallicity, age and
IMF. A summary is presented in \S4.

\section{INTEGRATED SYNTHETIC SPECTRA OF SINGLE-AGED STELLAR POPULATIONS}

\subsection{Synthetic Spectra of Individual Stars}

SB99 computed a grid of high resolution synthetic spectra in the wavelength range
$\lambda\lambda$ 6000 -- 10200 {\rm \AA}. The 
photospheric models by Kurucz (1992), Plez et al. (1992),
further unpublished models by Plez (1997), and Allard \& Hauschildt (1995)
were adopted.
The grid covers the following ranges of stellar parameters:
2500 $\leq$ T$_{\rm eff}$ $\leq$ 6000 K, 
-0.5 $\leq$ log g $\leq$ 5.0, [M/H] = -0.5, 0.0 and +0.5,
and [$\alpha$/Fe] = 0.0.
The molecular electronic systems included in the calculations are:
the CN (A$^2\Pi$--X$^2\Sigma$) red system,
the C$_2$ (A$^3\Pi_g$--X$^3\Pi_u$) Swan system,
the TiO $\gamma$(A$^3\Phi$--X$^3\Delta$), $\gamma'$(B$^3\Pi$--X$^3\Delta$),
$\delta$(b$^1\Pi$--a$^1\Delta$), $\epsilon$(E$^3\Pi$--X$^3\Delta$), and
$\phi$(b$^1\Pi$--d$^1\Sigma$) systems, and
the FeH (A$^4\Delta$--X$^4\Delta$), which includes the WFB.
The synthetic spectra were computed in steps of 0.02 {\rm \AA}
and rebinned to 1.0 {\rm \AA} with FWHM = 2.0 {\rm \AA}. 
More details on the atmospheric models and the atomic and
 molecular data employed are given in SB99. 

\subsection{Integrated Spectra of Single-aged Stellar Populations}
 
We have combined our grid of synthetic stellar spectra with isochrones
collected from the literature to build SSP synthetic spectra.  We
adopted the Padova isochrones (Bertelli et al. 1994) for $M >
0.6M_\odot$, and the isochrones from Baraffe et al. (1998) for $M \leq
0.6M_\odot$.  We have built integrated spectra of SSPs for [M/H] =
-0.5, 0.0 and +0.5, and two different IMF slopes: $x$ = 1.35 (Salpeter)
and $x$ = 2.0 (dwarf enriched). In Fig. \ref{Fig flux} we show examples
of the computed spectra in the wavelength range $\lambda\lambda$
6000-10200 {\rm \AA}, normalized at $\lambda$8157 {\rm \AA} and
convolved with a gaussian of FWHM = 9{\rm \AA}, for a range of ages
(Fig. \ref{Fig flux}a), metallicities (Fig. \ref{Fig flux}b) and IMFs
(Fig. \ref{Fig flux}c). As expected, NIR features are stronger in the
more metal-rich and older SSPs, since their spectra are dominated by
the cooler and redder stars.  With a higher contribution of M dwarfs to
the integrated light (Fig. \ref{Fig flux}c), the spectrum becomes
somewhat redder.

In order to test our SSP integrated spectra, we have
(1) compared results obtained with Padova and Geneva isochrones,
and (2) applied our calculations to the metal-rich
globular cluster NGC 6553, as described below.

\subsubsection{Geneva vs. Padova Isochrones}

A check of our results obtained with the Padova isochrones was carried
out by using the isochrones from the Geneva group (Schaller et al. 1992
as given in CD-ROM by Leitherer et al. 1996) for stars with $M >
0.6M_\odot$, to compute the integrated spectrum of a 13 Gyrs old SSP
with solar metallicity and Salpeter IMF.
 For lower masses we use Baraffe et al.'s isochrones in all cases. In
Figures
 \ref{isoch} and \ref{isocomp}  we compare, respectively, the Geneva
and Padova isochrones, and corresponding integrated spectra
 (normalized at $\lambda=8170{\rm\AA}$).

From Figure \ref{isoch} it can be seen that the Geneva and Padova isochrones
present differences in the red giant branch. While in the first
ascent giant branch the Geneva isochrones present a systematically
lower temperature ($\sim$100K), their red giant tip is much hotter and
brighter (for a given temperature) than in the Padova isochrones. The
coolest red giant stars in the Geneva isochrone have T$_{\rm
eff}\sim$3300K, while in the Padova isochrone red giants can be as cool
as T$_{\rm eff}\sim$2600K.

In spite of these large differences in the isochrones, the integrated
spectra look remarkably similar in Figure \ref{isocomp}. The integrated
spectrum computed from the Padova isochrones is brighter than the
Geneva one by less than 5\% in both ends of the interval. By looking at
Figure 11 of SB99, these small differences can be explained by the fact
that stars from the tip of the red giant branch (stars later than M4),
which are cooler in the Padova isochrones, dominate the integrated
light of old SSPs at $\lambda\simgreat8500{\rm\AA}$, while stars from
the warmer part of the red giant branch - RGB (spectral types from
mid-K to early-M), which are cooler in the Geneva isochrones, are
dominant in the bluer part of the spectral interval under study. The
differences in the studied indices are small ($<$ 10\% in equivalent
width) except for the Wing-Ford band, for which a 30\% difference in
equivalent width is found, due to the sensitivity of this feature to
the temperature of RGB tip stars.

The fact that we obtain so small spectral differences suggests the
existence of a mutual compensation between the above mentioned
differences in the isochrones. 

\subsubsection{The Metal-rich Globular Cluster NGC 6553}

Our choice to adopt the Padova set of
isochrones is based on our recent study of the T$_{\rm eff}$ scale of M
giants in the bulge metal-rich globular cluster NGC 6553 (SB99).
The metallicity of NGC 6553 found in the literature ranges
from -0.70 $<$ [Fe/H] $<$ +0.47 (see Table 5 in Barbuy et al. 1999).
High resolution spectroscopy tends to give lower values:
-0.7 $<$ [Fe/H] $<$ -0.2, whereas integrated photometry or spectroscopy
give higher values: -0.33 $<$ [Fe/H] $<$ +0.47.
Rutledge et al. (1997) gives [Fe/H] = -0.18 or -0.60 depending
whether they use the metallicity scale of globular clusters by
Zinn \& West (1984) or by Carretta \& Gratton (1997).
Barbuy et al. (1999) obtained [Fe/H] $\approx$ -0.55 and [$\alpha$/Fe]
$\approx$ +0.4, which would explain the discrepancy between results
obtained from high resolution measurements of Fe lines and integrated
spectra where lines of Fe and $\alpha$-elements give
together [Z/Z$_{\rm \odot}$] 
$\approx$ 0.0. Cohen et al. (1999) on the other hand, obtained
[Fe/H] $\approx$ -0.2 and [$\alpha$/Fe] $\approx$ 0.2.
In terms of our present purposes, since the spectra are dominated
by the M giants, whose spectra are in turn  dominated by TiO bands, and given
that both Ti and O are $\alpha$ elements, the use of an overall
metallicity of [M/H] $\approx$ 0.0 is in agreement with the above two
results (Barbuy et al. 1999 and Cohen et al. 1999).

 The T$_{\rm
eff}$-scale inferred by SB99 for NGC 6553, based on TiO bands, is more compatible
with the temperature values in the Padova isochrones,
which in particular have a more extended RGB tip;
the extension of the RGB of NGC 6553 can be seen in Bruzual et  al. (1997)
and Guarnieri et al. (1998). Moreover, in a study of the
T$_{\rm eff}$s of M giants from NGC 6528 (another metal rich globular
cluster from the Galactic bulge, see Ortolani et al. 1995), 
we determined T$_{\rm eff}$ = 3000 K for
a bonafide M giant of this cluster, which is 300K cooler than the
lowest T$_{\rm eff}$ predicted by the Geneva isochrone (Schiavon et
al. 1999, in preparation).

\section{NaI DOUBLET, CaII TRIPLET, FeH WING-FORD BAND
AND TiO BANDS IN SINGLE STELLAR POPULATIONS}

In this Section we discuss the effects of metallicity and IMF upon NIR
spectral indices commonly used for stellar population diagnosis in
integrated spectra of galaxies and clusters (e.g. Cohen 1978, 1979;
Delisle \& Hardy 1992). The dependence of the TiO bands, NaI doublet,
CaII triplet and the WFB on stellar atmospheric parameters have been
studied in previous papers (Erdelyi-Mendes \& Barbuy 1991; Milone \&
Barbuy 1994; Schiavon et al. 1997a,b; SB99).

In Table 1 we report the definition of the spectral indices used in
this work, including TiO $\lambda$6600 {\rm \AA}, NaI $\lambda$8190
{\rm \AA}, CaII $\lambda$8662 {\rm \AA} and the FeH WFB at
$\lambda$9900 {\rm \AA}, as given in the literature.  In Table 2 we
list the equivalent width of the indices measured on  the spectra of
13 Gyr SSPs, as a function of metallicity, IMF and age, where  convolutions 
with gaussian profiles of FWHM = 9 and 25{\rm
\AA} were adopted.  We note that two definitions of the NaI index are
used: Na$_{\rm FF}$ and Na$_{\rm S97}$ as defined in Faber \& French
(1980, hereafter FF) and Schiavon et al. (1997a, hereafter S97a)
respectively.

\subsection{CaII Triplet, TiO Bands and FeH WFB}

Figures \ref{feeimf}a-f show synthetic spectra of 13 Gyr SSPs in the
regions of the TiO band ($\lambda$6600 \AA), the WFB ($\lambda$9900
\AA), and CaII ($\lambda$8662 \AA).  Figs. \ref{feeimf}(a,c,e)
correspond to the [M/H] = -0.5, 0.0 and +0.5 models for the Salpeter
IMF, and Figs. \ref{feeimf}(b,d,f) to the [M/H] = -0.5 model for the
Salpeter and the $x$ = 2.0 IMF.

Figs. \ref{feeimf}(a,c,e) show that the TiO bands are the most
metallicity-sensitive feature under analysis.  We restrict our discussion to the
TiO$_{\rm 6600}$ band, since all bands of the same molecule have essentially
the same behavior. 

The WFB and the CaII triplet are also very sensitive to metallicity and
weakly sensitive to IMF variations.  It has been suggested in the
literature  (Couture \& Hardy 1993 and references therein) that the WFB
and the CaII triplet (Delisle \& Hardy 1992; Jones et al. 1984) are
sensitive to the contribution of M dwarfs to the integrated spectrum.
Our computations do not confirm this.  Fig. \ref{feeimf}b shows a
residual sensitivity of the TiO$_{6600}$ index to IMF variations, while
the WFB (Figs. \ref{feeimf}c,d) is only weakly IMF-sensitive, but its
response to metallicity is much stronger, being due to the
contamination by TiO lines from the (2,3) vibrational band of the
$\delta$ system.

\subsection{The NaI Doublet}

Figures \ref{naI}a,b show  spectra of 13 Gyr SSPs  in the
region of the NaI $\lambda$8190 \AA\ feature.
Fig. \ref{naI}a correspond to the [M/H] = -0.5, 0.0 and +0.5 models for the
Salpeter IMF, and Fig. \ref{naI}b to the [M/H] = -0.5 model for the Salpeter
and the $x$ = 2.0 IMF. 

Figs. \ref{naI}a,b show that the NaI doublet is by far the most
IMF-sensitive index studied here. The Na$_{\rm S97}$ index displays a
strong dependence on IMF (Table 2).  It also shows a dependence on
metallicity.  Besides, the NaI doublet is contaminated by TiO lines
(which are very sensitive to metallicity) at the red side of the
feature.

It has been suggested in the literature (Cohen 1978; Xu et al. 1989;
Alloin \& Bica 1989; Terndrup et al. 1990; Delisle \& Hardy 1992) that
the NaI doublet is 
more sensitive to metallicity than to IMF variations.
  Figs. \ref{naI} a,c show that the NaI {\it lines} are indeed
strongly sensitive to metallicity.  However, the NaI {\it index}, as
defined in the literature (Table 1), has its metallicity-sensitivity
reduced by the TiO lines that contaminate the red continuum window,
which is thus lowered in the spectra of metal-rich stellar populations.

Figs. \ref{naI} and the values in Table 2 suggest that the NaI index is
very sensitive to IMF. In Fig. \ref{indices1} the equivalent width of
 NaI, according to the definitions by FF and Schiavon et al. (1997) and
CaII lines (normalized by the values corresponding to a SSP of 13Gyrs,
solar metallicity and Salpeter IMF), are plotted against the exponent
of a power-law IMF, for different metallicities and an age of 13Gyrs.
The NaI index, in both definitions, is as sensitive to IMF as to
metallicity. We therefore conclude that the possibility that an
enhancement of the NaI lines may be due to a dwarf-enriched IMF cannot
be excluded.  An uncertainty in metallicity of $\Delta$[M/H] = 0.25 dex
translates into an uncertainty of 15\% in x 
for x $>$ 1, and 50\% for x $<$ 1;
therefore the use of  the NaI doublet as an IMF indicator requires
narrow constraints on the average metallicity of the stellar
population.

Figs. \ref{naI} suggest that in the integrated spectra of galaxies, where lines
are broadened by stellar velocity dispersion, IMF and metallicity
effects may be distinguished by their different influence on the line
shape, as proposed by Boroson \& Thompson (1991). A higher metallicity
leads to a profile with a deeper red side, while a profile with a
deeper blue side can be due to either a stronger metallicity or a
dwarf-enriched IMF.  
%In other words, the possibility that an
%enhancement of the NaI lines may be %due to a dwarf-enriched IMF cannot
%be excluded.

\subsection{Sensitivity of the Features to Age}

In Table 3 are given computed synthetic indices
convolved with FWHM = 25{\rm \AA} for single
stellar populations for ages 3 $<$ Age (Gyr) $<$ 17 and
-0.5 $<$ [M/H] $<$ +0.5 for a Salpeter IMF.

In Fig. \ref{indices2} we plot the four indices  versus the age of
SSPs for [M/H] = -0.5 and 0.0. It appears that the TiO band, the
CaII triplet, and the WFB are essentially insensitive to age.
Note the strong metallicity dependence of the
TiO bands.   The NaI
doublet is very sensitive to age, as well as to the IMF slope and
metallicity, which makes the behavior of this feature quite complex.
It is of interest to explore indices such as TiO bands and the CaII
triplet to help disentangle the age-metallicity degeneracy in composite
stellar populations.

\section{SUMMARY}

We have built high resolution synthetic spectra of SSPs in the
wavelength range $\lambda\lambda$ 6000-10200 {\rm \AA} for SSPs of
[M/H] = -0.5, 0.0 and +0.5, and IMF index $x$ = 0.1-2.0.  The basic
ingredients for such SSP models are the isochrones by Bertelli et al.
(1994) and Baraffe et al. (1998), combined to a
 grid of synthetic spectra computed for a wide range of stellar
parameters (SB99).

In order to make our computations useful for observers, we
present a table of TiO $\lambda$6600 {\rm \AA}, NaI $\lambda$8190
{\rm \AA}, CaII $\lambda$8662 {\rm \AA} and the FeH WFB at
$\lambda$9900 {\rm \AA} 
 indices in single stellar populations, as a function of
metallicity, IMF and age.

The main conclusions of this paper are:

1) We have shown that the TiO bands, the CaII triplet, and the WFB are very
sensitive to metallicity and essentially invariant with IMF and age.
2) The NaI doublet is far more intense for a dwarf-enriched than for
the Salpeter IMF, and the feature is also affected by metallicity
and age variations.
3) The better understanding of the NIR indices as a function
of metallicity, age and IMF obtained through the present spectrum
synthesis computations, provides us with tools, which combined
to optical indices, can help disentangle the age - metallicity - 
IMF degeneracy in composite stellar systems.

\acknowledgements
We are grateful to the referee for helpful comments on an earlier
version of this paper. Calculations were carried out in a DEC-Alpha
3000/700 workstation. RPS acknowledges support from CNPq fellowship
No. 3000098/4.

\begin{deluxetable}{lccccccccc}
\tablenum{1}
\tablewidth{0pt}
\tablecaption{Definition of spectral indices measured}
\tablehead{
\colhead {Index} &
\colhead { Blue continuum} &
\colhead { Bandpass} &
\colhead { Red continuum} }
\startdata
EW$_{6600}$ & 6512.1-6538.1 & 6617.2-6992.5 & 
7036.9-7048.0\nl
EW$_{8190}^{\rm FF}$ & 8169.0-8171.0 & 8172.0-8209.0 & 8209.0-8211.0 \nl
EW$_{8190}^{\rm S97}$ & 8171.5-8172 & 8172-8197 & 8233.5-8234.2 \nl
EW$_{\rm Ca8662}$ & 8637.2-8646.2 & 8653.2-8668.4 & 8847.6-8854.0 \nl
EW$_{\rm WFB}$ & 9891.8-9895.1 & 9895.1-9958.6    & 9958.6-9962.2\nl
\enddata
\end{deluxetable}

\vfill\eject

\begin{deluxetable}{lccccccccc}
\tablenum{2}
\tablewidth{0pt}
\tablecaption{Spectral indices
(equivalent widths in {\rm \AA}) in synthetic Single Stellar Populations
of 13 Gyr, for convolutions of 9 and 25 {\rm \AA}}
\tablehead{
\colhead {[M/H]} &
\colhead {$x$} &
\colhead {Na$_{\rm FF}$ } &
\colhead {Na$_{\rm S97}$ } &
\colhead {TiO$_{6600}$ } &
\colhead {Ca$_{8662}$ } &
\colhead {WFB}}
\startdata
      &      &      & FWHM=9{\rm \AA} &  &  &  \nl  
--0.5 & 1.35 & 1.03 & 0.95 & 9.3 & 1.41 & 0.56 \nl  
  0.0 & 1.35 & 1.15 & 1.19 & 24.9 & 1.57 & 1.35 \nl
 +0.5 & 1.35 & 1.45 & 1.48 & 34.0 & 1.80 & 2.03 \nl
--0.5 & 2.0  & 1.56 & 1.40 & 11.2 & 1.41 & 0.72 \nl
  0.0 & 2.0  & 1.78 & 1.70 & 28.5 & 1.50 & 1.52 \nl
      &      &      & FWHM=25{\rm \AA} &  &  &  \nl  
--0.5 & 1.35 & 0.47 & 0.55 & 7.48 & 0.52 & 0.28 \nl  
  0.0 & 1.35 & 0.51 & 0.69 & 21.7 & 0.78 & 0.72 \nl
 +0.5 & 1.35 & 0.67 & 0.88 & 29.7 & 0.90 & 1.03 \nl
--0.5 & 2.0  & 0.80 & 0.81 & 9.38 & 0.60 & 0.45 \nl
  0.0 & 2.0  & 0.90 & 0.99 & 25.3 & 0.69 & 0.90 \nl

\enddata
\end{deluxetable}

\vfill\eject

\begin{deluxetable}{lccccccccc}
\tablenum{3}
\tablewidth{0pt}
\tablecaption{Spectral indices
(equivalent widths in {\rm \AA} in synthetic Single Stellar Populations as a 
function of age and metallicity, for a Salpeter IMF, adopting
a convolution of FWHM = 25}
\tablehead{
\colhead {age} &
\colhead {[M/H]} &
\colhead {Na$_{\rm FF}$ } &
\colhead {Na$_{\rm S97}$ } &
\colhead {TiO$_{6600}$ } &
\colhead {Ca$_{8662}$ } &
\colhead {WFB}}
\startdata
  3   &-0.5& 0.532    &   0.578    &    0.379    &	 0.749	   &0.339 \nl
  5   &-0.5& 0.634    &   0.633    &    0.377    &	 0.644	   &0.381 \nl
  6   &-0.5& 0.676    &   0.663    &    0.387    &	 0.689	   &0.393 \nl
  8   &-0.5& 0.723    &   0.691    &    0.397    &	 0.691	   &0.391 \nl
 10   &-0.5& 0.799    &   0.736    &    0.385    &	 0.599	   &0.411 \nl
 12   &-0.5& 0.886    &   0.773    &    0.326    &	 0.717	   &0.379 \nl
 13   &-0.5& 0.925    &   0.796    &    0.345    &	 0.663	   &0.394 \nl
 15   &-0.5& 0.992    &   0.832    &    0.341    &	 0.748	   &0.407 \nl
 17   &-0.5& 1.073    &   0.878    &    0.319    &	 0.640	   &0.401 \nl
  3   & 0.0& 0.422    &   0.660    &    0.809    &	 0.926	   &0.884 \nl
  5   & 0.0& 0.558    &   0.739    &    0.864    &	 0.993	   &0.886 \nl
  6   & 0.0& 0.647    &   0.788    &    0.893    &	 1.037	   &0.961 \nl
  8   & 0.0& 0.698    &   0.828    &    0.928    &	 0.955	   &0.975 \nl
 10   & 0.0& 0.784    &   0.879    &    0.971    &	 0.946	   &1.020 \nl
 12   & 0.0& 0.888    &   0.940    &    0.970    &	 0.908	   &1.019 \nl
 13   & 0.0& 1.000    &   1.000    &    1.000    &	 1.000	   &1.000 \nl
 15   & 0.0& 1.077    &   1.034    &    0.945    &	 0.891	   &0.957 \nl
 17   & 0.0& 1.071    &   1.043    &    1.058    &	 1.016	   &1.010 \nl
  3   & 0.5& 0.530    &   0.813    &    0.892    &	 1.178	   &1.005 \nl
  5   & 0.5& 0.673    &   0.900    &    0.994    &	 1.174	   &1.163 \nl
  6   & 0.5& 0.775    &   0.966    &    1.128    &	 1.193	   &1.264 \nl
  8   & 0.5& 0.912    &   1.038    &    1.139    &	 1.276	   &1.246 \nl
 10   & 0.5& 0.973    &   1.078    &    1.237    &	 1.080	   &1.280 \nl
 12   & 0.5& 1.220    &   1.214    &    1.303    &	 0.895	   &1.364 \nl
 13   & 0.5& 1.315    &   1.272    &    1.367    &	 1.150	   &1.436 \nl
 15   & 0.5& 1.353    &   1.299    &    1.467    &	 1.164	   &1.514 \nl
 17   & 0.5& 1.406    &   1.325    &    1.403    &	 1.051	   &1.353 \nl

\enddata
\end{deluxetable}

\clearpage

\centerline {\bf FIGURE CAPTIONS}

\figcaption{Synthetic spectra 
in the wavelength range $\lambda\lambda$ 6000 - 10200 {\rm \AA} for 
(a) Salpeter IMF, [M/H] = 0.0 and ages of 3 Gyr (solid line),
8 Gyr (dashed line) and 13 Gyr (dotted line);
(b) Salpeter IMF and [M/H] = -0.5 (solid line), 0.0 (dashed line)
 and +0.5 (dotted line), 
normalized at $\lambda$8170 {\rm\AA};
(c) [M/H] and IMF index $x$ = 1.35 (solid line) and 2.0 (dashed line). 
\label{Fig flux}}

\figcaption{ Isochrones of 13Gyr and [Z/Z$_{\rm \odot}$] = 0.0 from
the  Padova  (open circles) and Geneva groups (filled squares).
 \label{isoch}}

\figcaption{Integrated spectra generated with the use of Geneva
(solid line) and Padova (dotted line)
 isochrones  (normalized at $\lambda=8170{\rm\AA}$)
are compared.                            
 \label{isocomp}}

\figcaption{ Synthetic spectra for single stellar
populations with 13$Gyr$, a Salpeter IMF and [M/H] = -0.5 (solid), 0.0 (dashed) and +0.5 
(dotted) for the features 
(a) TiO$_{6600}$, (c) WFB, (e) CaII triplet;
and synthetic spectra for SSPs of [M/H] = -0.5
for Salpeter (solid) and dwarf-enriched (dashed) IMFs for
(b) the TiO$_{6600}$, (d) WFB, (f) CaII triplet.
 \label{feeimf}}

\figcaption{ Synthetic spectra of the NaI feature for SSPs of 13 Gyr,
Bertelli et al. (1994) isochrones, and
(a) Salpeter IMF and [M/H] = -0.5 (solid), 0.0 (dashed) and +0.5 (dotted);
(b) [M/H] = -0.5 and IMF index $x$ = 1.35 (solid) and 2.0 (dashed).
 \label{naI}}

\figcaption{
CaII, NaI$_{\rm S97}$ and NaI$_{\rm FF}$ vs. IMF index $x$, for
[M/H] = -0.5, 0.0 and +0.5 and age=13Gyrs. 
\label{indices1}}

\figcaption{Indices for TiO$_{6600}$, WFB, NaI and CaII
measured on the
spectra of single stellar populations as a function of age (Gyrs)
for [M/H] = -0.5 (solid lines) and 0.0 (dotted lines). Equivalent
widths are normalized to the value correpsonding to a SSP with solar metallicity
13Gyrs and a Salpeter IMF.
\label{indices2}}

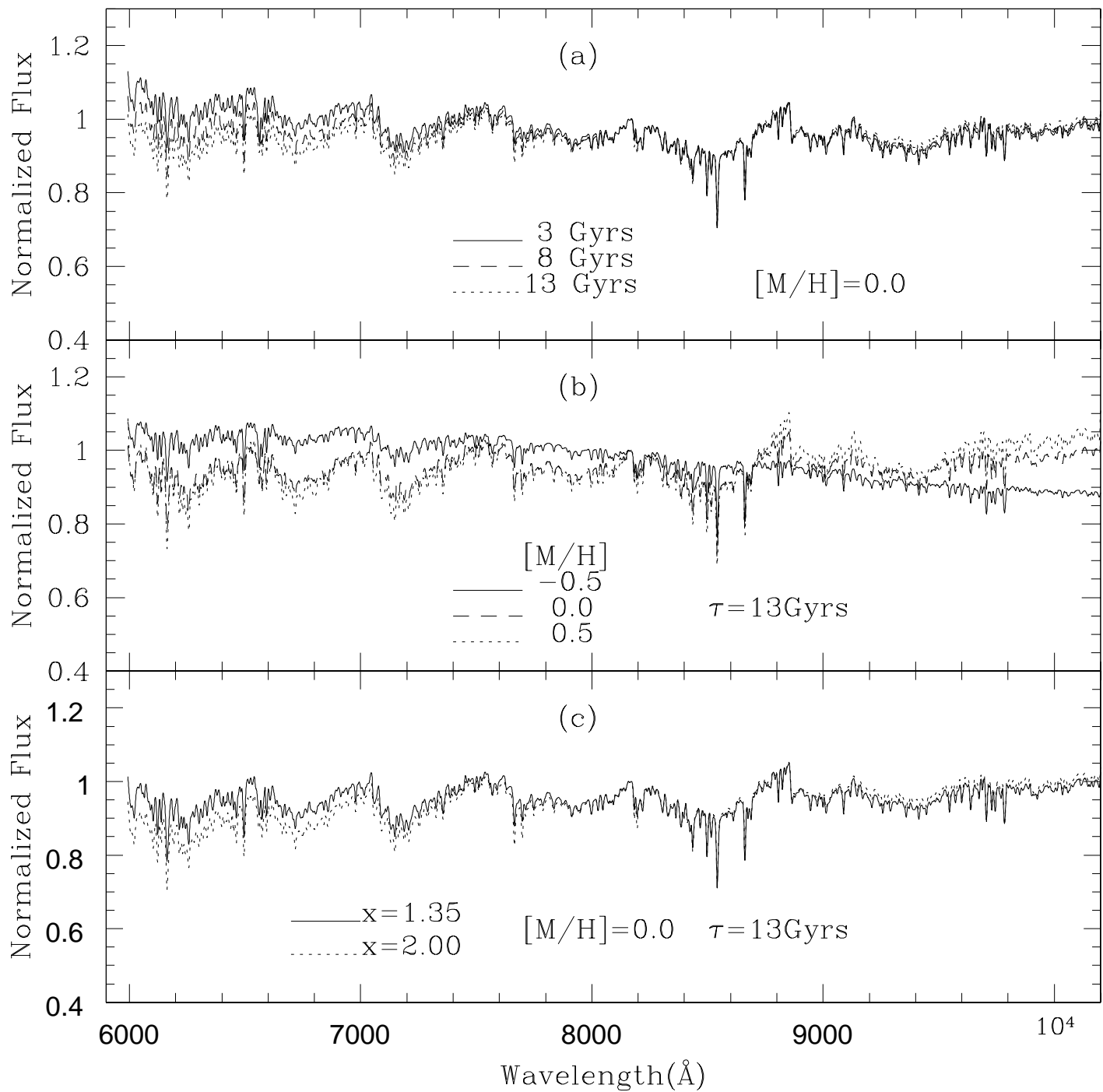
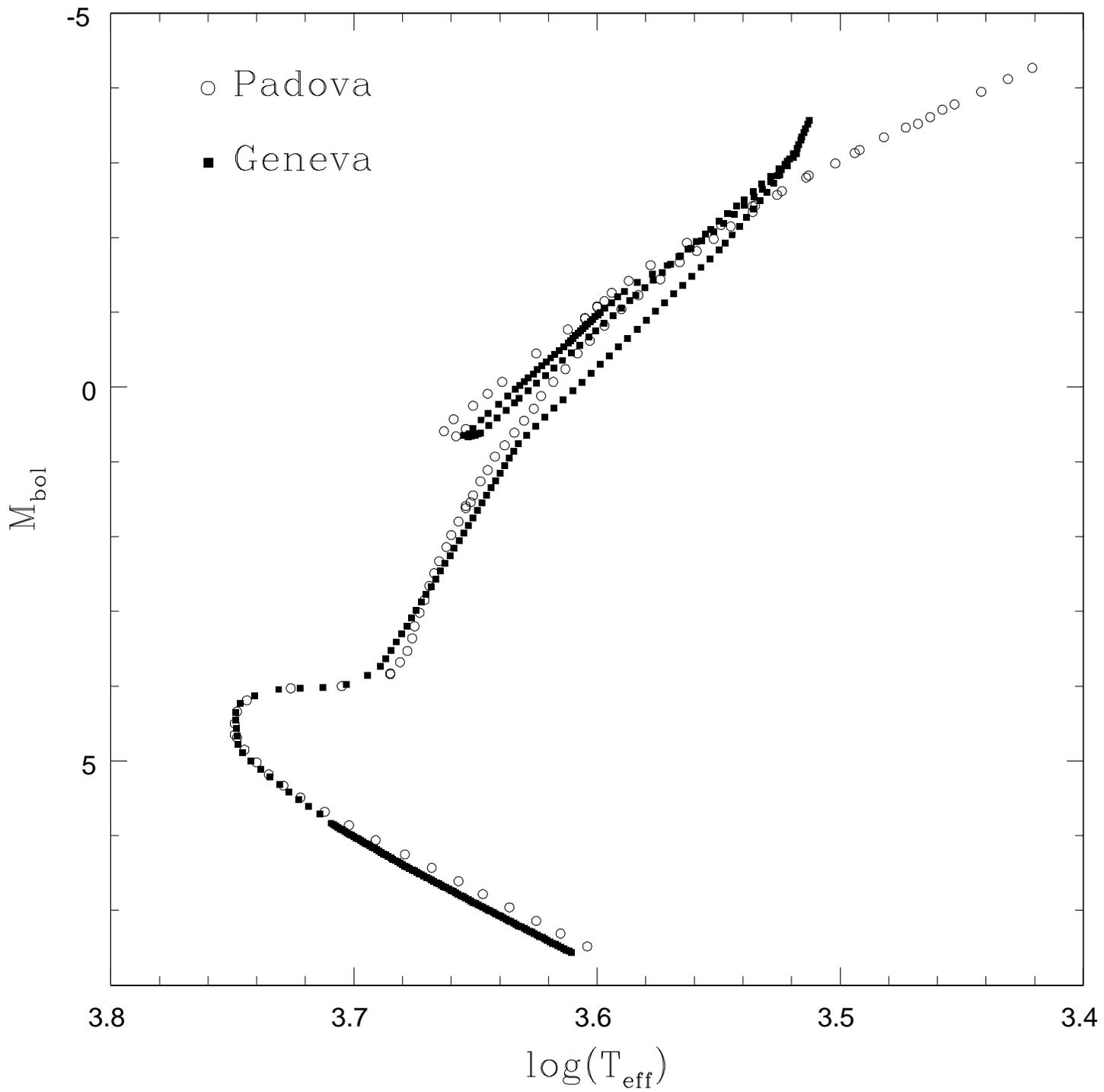
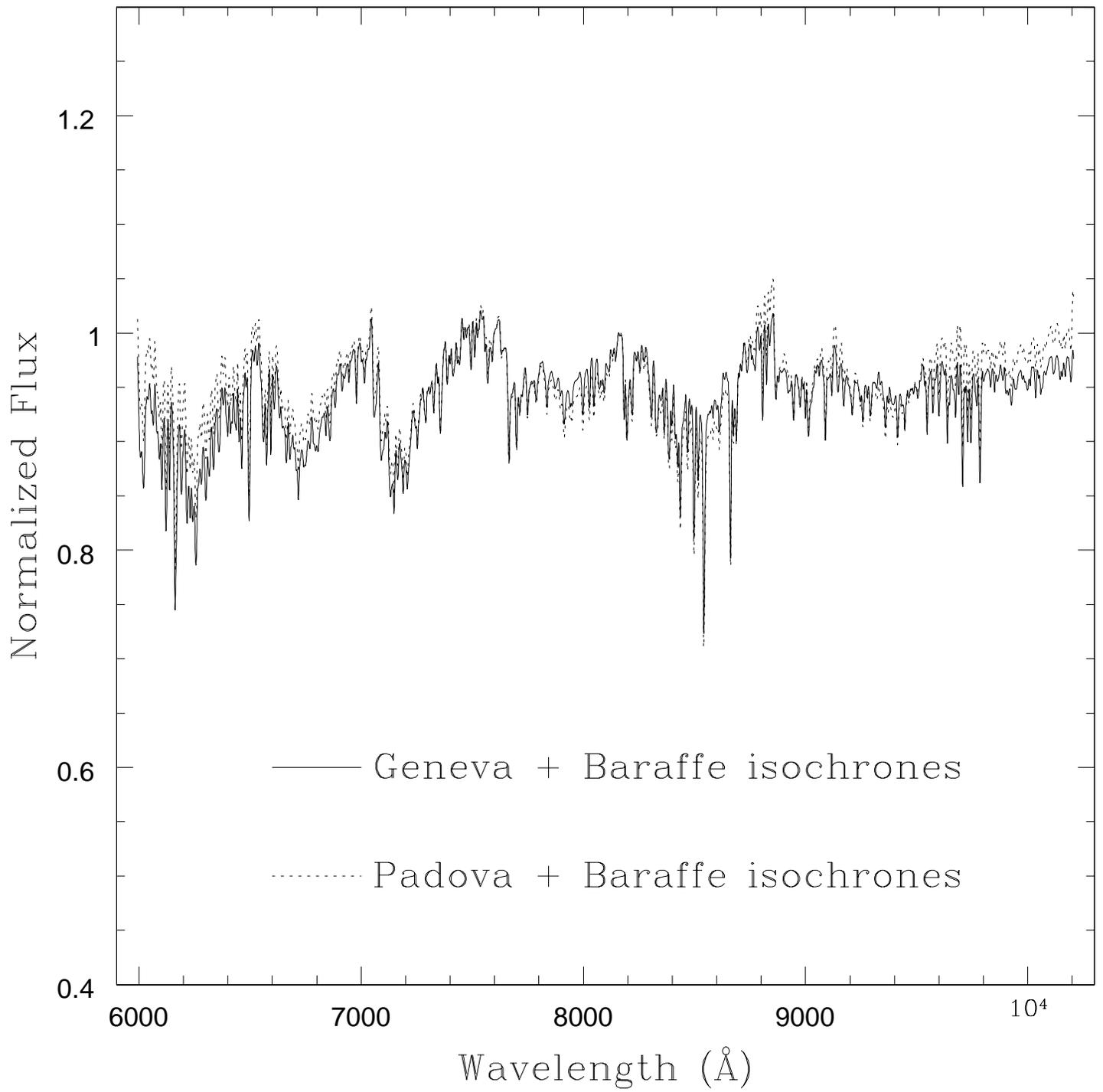
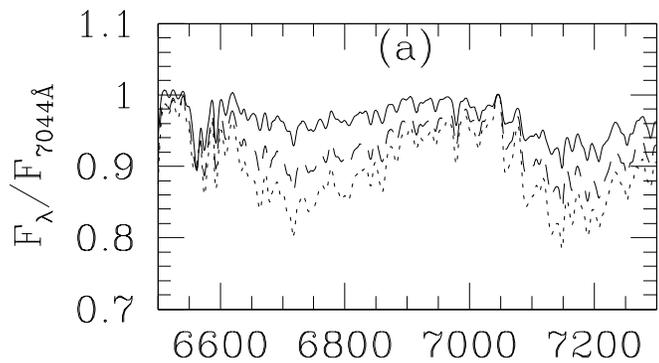
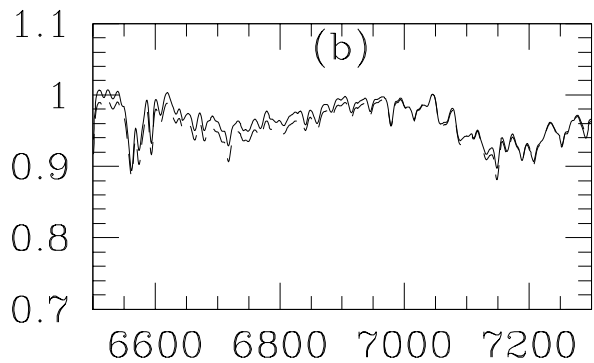
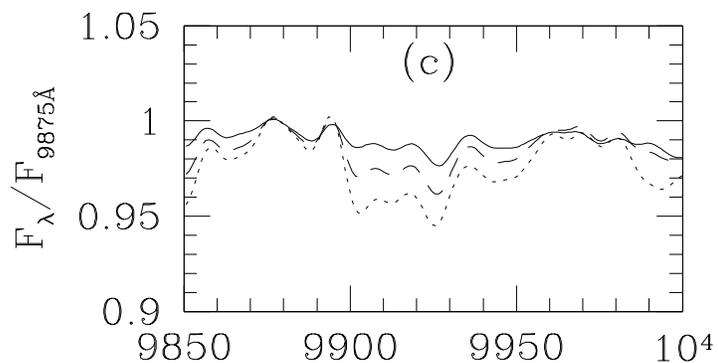
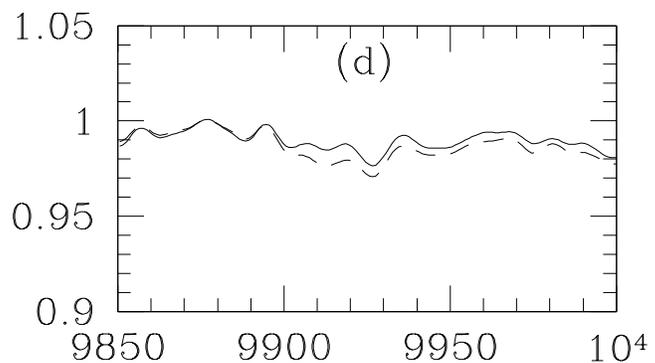
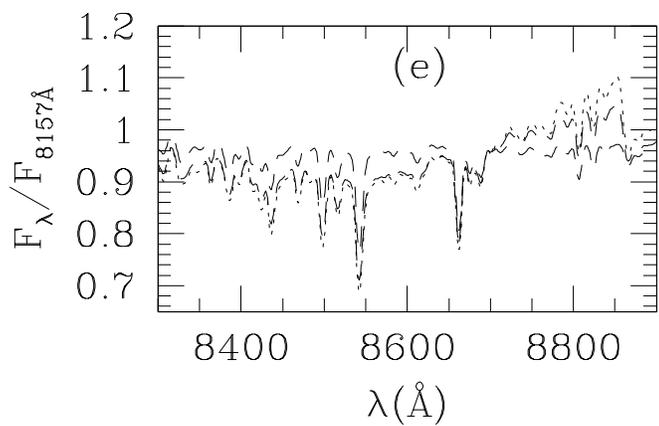
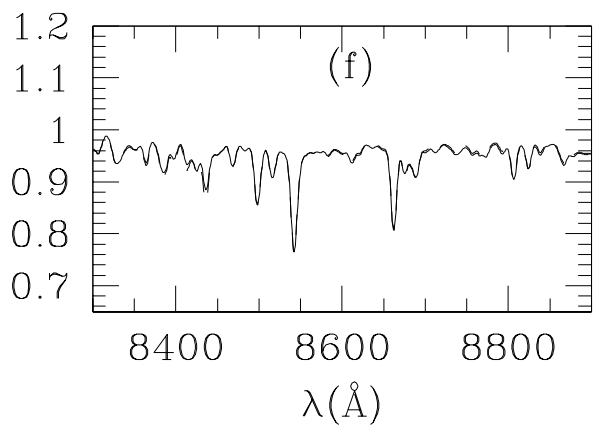
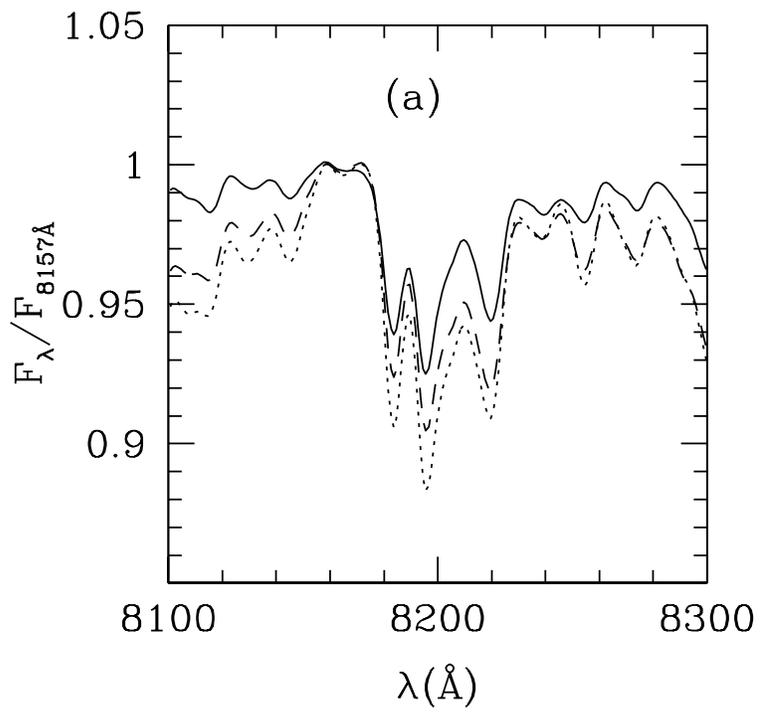
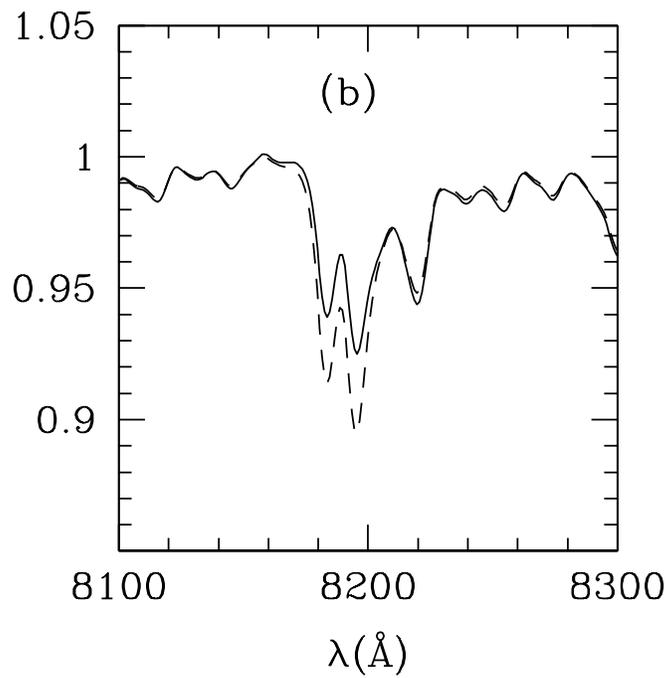
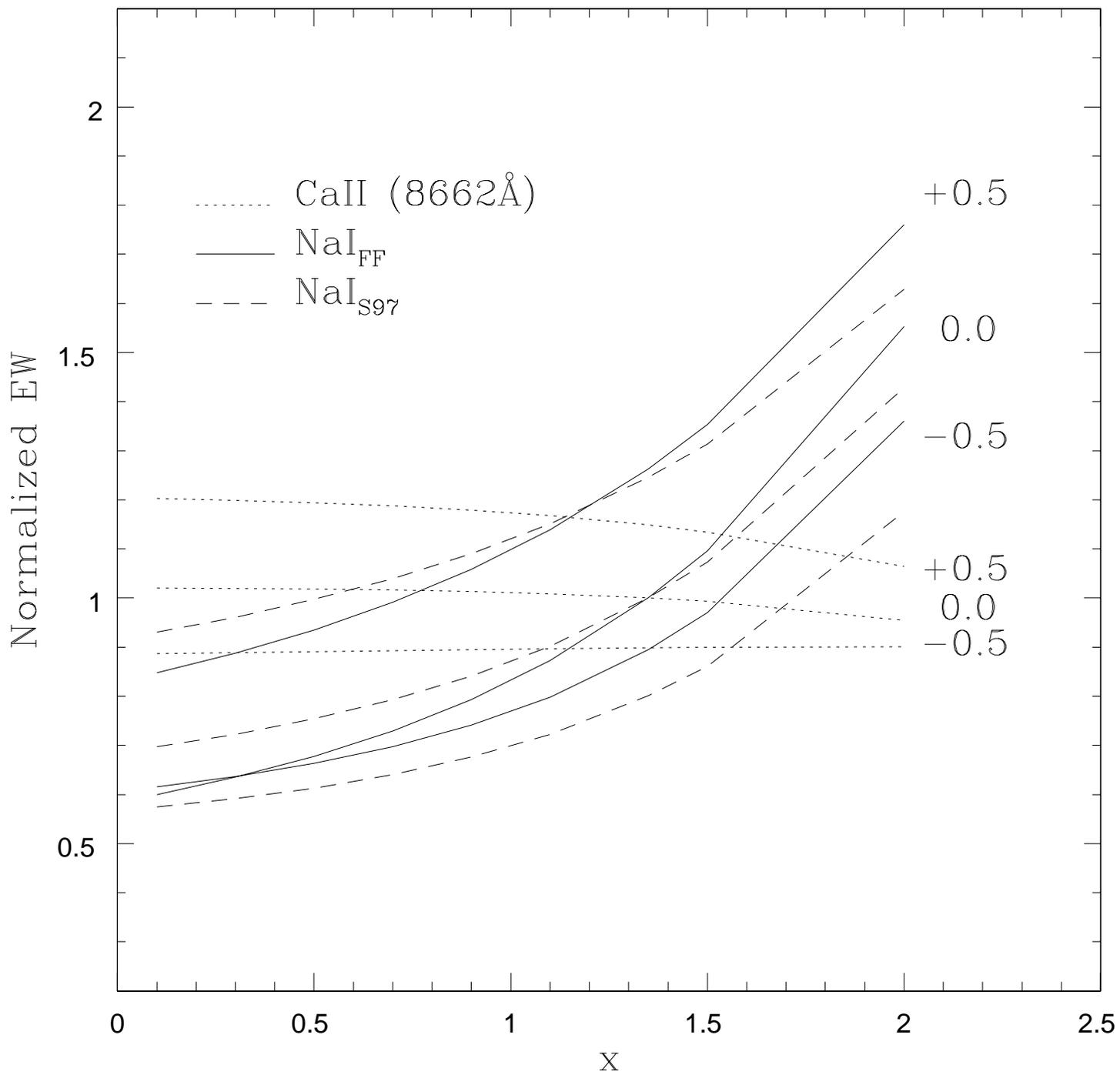
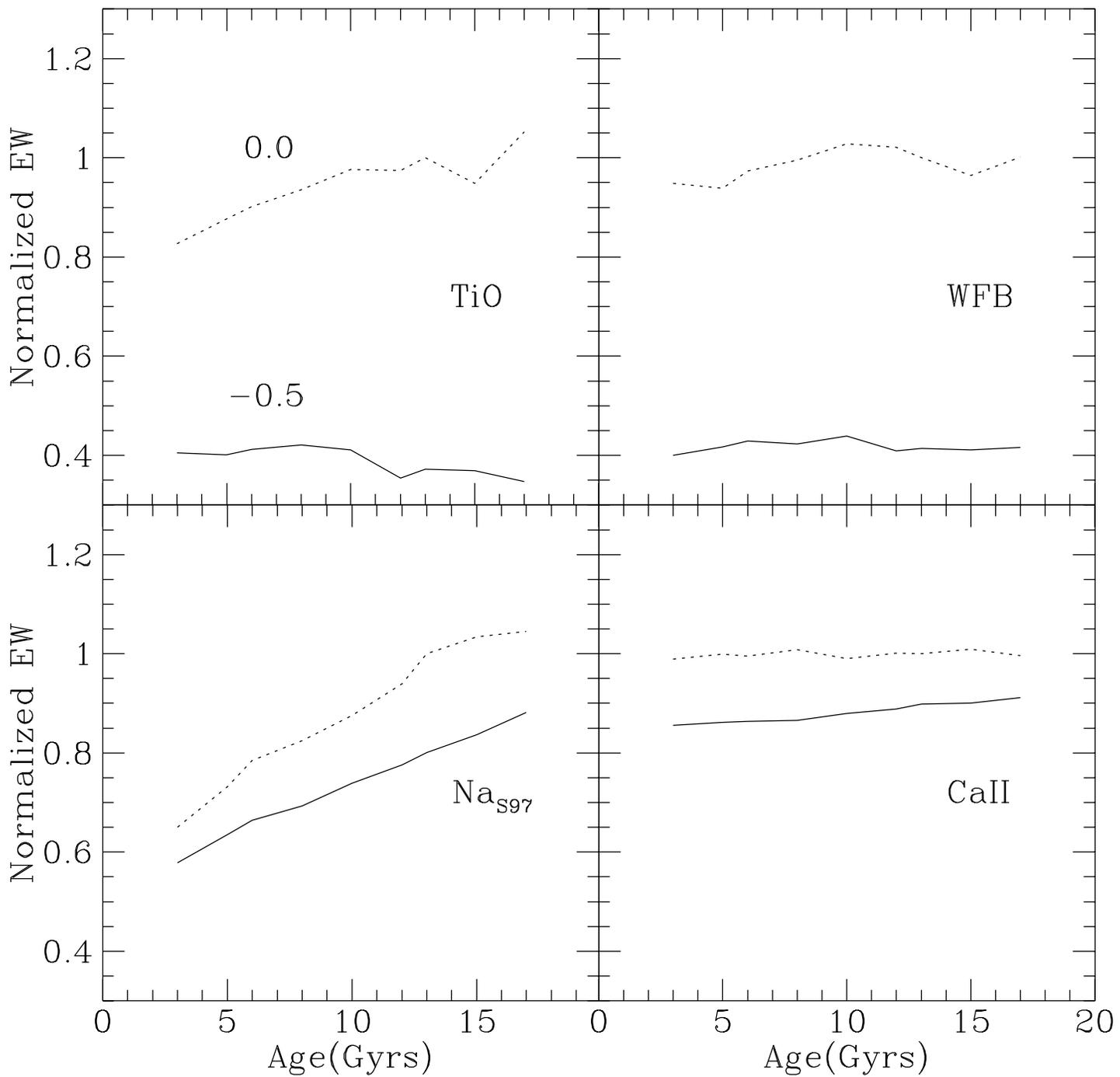
\end{document}